\documentclass[aps,tightlines]{revtex4}

\usepackage{amsfonts}
\usepackage{amsmath}
\usepackage{amssymb}
\usepackage{graphicx}

\begin{document}

\title[Local Unitary...]{Local Unitary Invariants for N-qubit Pure State}
\author{S. Shelly Sharma}
\email{shelly@uel.br}
\affiliation{Depto. de F\'{\i}sica, Universidade Estadual de Londrina, Londrina
86051-990, PR Brazil }
\author{N. K. Sharma}
\email{nsharma@uel.br}
\affiliation{Depto. de Matem\'{a}tica, Universidade Estadual de Londrina, Londrina
86051-990 PR, Brazil }
\thanks{}

\begin{abstract}
The concept of negativity font, a basic unit of multipartite entanglement,
is introduced. Transformation properties of determinants of negativity fonts
under local unitary (LU) transformations are exploited to obtain relevant N
qubit polynomial invariants and construct entanglement monotones, from first
principles. It is shown that entanglement monotones that detect the
entanglement of specific parts of the composite system may be constructed to
distinguish between states with distinct types of entanglement. The
structural difference between entanglement monotones for odd and even number
of qubits is brought out.
\end{abstract}

\maketitle

%\pacs{03.67.Mn, 03.65.Ud, 03.67.-a}

In 1935 Schr\"{o}dinger \cite{schr35} coined the term `entanglement' to
describe quantum correlations that make it possible to alter the properties
of a distant system instantaneously by acting on a local system. A spin
singlet is an example of entangled state of two spin half particles. A qubit
is any two level quantum system with basis states represented by $\left\vert
i\right\rangle $, $i=0$ and $1$. The spin singlet is an entangled state of
two qubits. For a pure state of bipartite quantum system consisting of two
distinguishable parts $A$ and $B$, each of arbitrary dimension, negativity 
\cite{zycz98} of partially transposed state operator \cite{pere96} is known
to be an entanglement monotone \cite{vida02}. How properties of one part of
a multipartite quantum system are altered by local operations on other parts
at distinct remote locations is a complex question. In this letter, we
present a novel approach to construct meaningful LU invariants for
multi-qubit systems from first principles that is by examining the effect of
local unitaries on different parts of the composite system. Our method,
illustrated for four qubit case in ref. \cite{shar110} introduces basic
units of entanglement, referred to as negativity fonts. A negativity font is
defined as a two by two matrix of probability amplitudes that determines the
negative eigen values of a specific four by four submatrix of a partially
transposed state operator. It was shown earlier \cite{shar108} that a
partial transpose can be written as a sum of $K-$way ($2\leq K\leq N$)
partial transposes. A $K-$way partial transpose contains information about $%
K\ $body corelations of multipartite system. Contibutions of partial
transposes to global negativity, referred to as partial $K-$way negativities
are not unitary invariants, but when calculated for canonical states for
three qubits \cite{shar07, shar208} and four qubit \cite{shar09} coincide
with entanglement monotones. This article complements our earlier work by
outlining a direct method to obtain multiqubit invariants relevant to the
construction of entanglement monotones without reaching the canonical state
to calculate partial $K-$way negativities. Multi qubit unitary invariants
are obtained by examining the transformation properties of negativity fonts
present in global partial transpose \cite{pere96} and $K-$way ($2\leq K\leq
N $) partially transposed matrices \cite{shar09} constructed from $N-$qubit
state operator. The mathematical form of resulting multiqubit invariants for
a given state reveals the entanglement microstructure of the state.

Multi qubit invariants, written in terms of determinants of negativity
fonts, are essentially relations between intrinsic negative eigenvalues of
selected $4\times 4$ submatrices of $K-$way partially transposed matrices.
In the case of four qubits, the standard approach from invariant theory has
lead to the construction of a complete set of SL-invariants \cite%
{luqu03,luqu06} and algorithm for constructing N-qubit invariants is given
in ref. \cite{luqu07}. Results for five qubits have also been reported \cite%
{luqu06}. The N qubit invariants for even number of qubits have been
reported earlier in \cite{wong01} and for even and odd number of qubis in 
\cite{li07}. Focus is on geometric aspects of such invariants in Refs. \cite%
{leva04,leva05,leva06}. Independent of these approaches, a method based on
expectation values of antilinear operators with emphasis on permutation
invariance of the global entanglement measure \cite{oste05,oste06}, has been
suggested. The number of polynomial invariants is known to increase very
fast with number of qubit. However, in general, a small number of invariants
is needed to qualify and quantify the entanglement. The advantage of our
approach is that it is easily applied to obtain relevant invariants for any
state at hand not necessarily the general state or canonical state. Our
results bring out the structural difference between LU invariants for $N-$%
odd and $N-$even qubits through the nature of $K-$way negativity fonts
present in respective invariants. For multipartite case, one needs
in-equivalent entanglement measures \cite{coff00,dur00,wong01}. To show that
the method can be used to construct entanglement monotones that detect the
entanglement of specific parts of the composite system, four qubit
invariants to detect entanglement of a pair of qubits due to four-way
correlations are obtained.

The entanglement of qubits $A_{1}$ and $A_{2}$ in pure state $\widehat{\rho }%
^{A_{1}A_{2}}=\left\vert \Psi ^{A_{1}A_{2}}\right\rangle \left\langle \Psi
^{A_{1}A_{2}}\right\vert $ where 
\begin{equation}
\left\vert \Psi ^{A_{1}A_{2}}\right\rangle =a_{00}\left\vert 00\right\rangle
_{^{A_{1}A_{2}}}+a_{10}\left\vert 10\right\rangle
_{^{A_{1}A_{2}}}+a_{01}\left\vert 01\right\rangle
_{^{A_{1}A_{2}}}+a_{11}\left\vert 11\right\rangle _{A_{1}A_{2}},
\label{twoqubit}
\end{equation}%
is measured by the negativity of four by four matrix $\left( \rho
^{A_{1}A_{2}}\right) _{G}^{T_{1}}$ obtained by partially transposing the
state of qubit $A_{1}$ in $\widehat{\rho }^{A_{1}A_{2}}$. We refer to two by
two matrix $\nu ^{00}=\left[ 
\begin{array}{cc}
a_{00} & a_{01} \\ 
a_{10} & a_{11}%
\end{array}%
\right] $ as a negativity font of $\left( \rho ^{A_{1}A_{2}}\right)
_{G}^{T_{1}}$. The squared negativity of $\left( \rho ^{A_{1}A_{2}}\right)
_{G}^{T_{1}},$ is given by $\left( N_{G}^{A_{1}}\right) ^{2}=4\left\vert
\det \nu ^{00}\right\vert ^{2}$. If $\det \nu ^{00}=0$, the state is
separable. A general N-qubit pure state reads as%
\begin{equation}
\left\vert \Psi ^{A_{1},A_{2},...A_{N}}\right\rangle
=\sum_{i_{1}i_{2}...i_{N}}a_{i_{1}i_{2}...i_{N}}\left\vert
i_{1}i_{2}...i_{N}\right\rangle ,  \label{nstate}
\end{equation}%
where $\left\vert i_{1}i_{2}...i_{N}\right\rangle $ are the basis vectors
spanning $2^{N}$ dimensional Hilbert space, and $A_{p}$ is the location of
qubit $p$. The coefficients $a_{i_{1}i_{2}...i_{N}}$ are complex numbers.
The basis states of a single qubit are labelled by $i_{m}=0$ and $1,$ where $%
m=1,...,N$. The global partial transpose of $N$ qubit state $\widehat{\rho }%
=\left\vert \Psi ^{A_{1}A_{2}...A_{N}}\right\rangle \left\langle \Psi
^{A_{1}A_{2}...A_{N}}\right\vert $ with respect to qubit $p$ is constructed
from the matrix elements of $\widehat{\rho }$ through 
\begin{equation}
\left\langle i_{1}i_{2}...i_{N}\right\vert \widehat{\rho }%
_{G}^{T_{p}}\left\vert j_{1}j_{2}...j_{N}\right\rangle =\left\langle
i_{1}i_{2}...i_{p-1}j_{p}i_{p+1}...i_{N}\right\vert \widehat{\rho }%
\left\vert j_{1}j_{2}...j_{p-1}i_{p}j_{p+1}...j_{N}\right\rangle .
\end{equation}%
If $\widehat{\rho }$ is a pure state, then the negative eigenvalue of $%
4\times 4$ sub-matrix of $\widehat{\rho }_{G}^{T_{p}}$ in the space spanned
by distinct basis vectors $\left\vert
i_{1}i_{2}...i_{p}...i_{N}\right\rangle $, $\left\vert
j_{1}j_{2}...j_{p}...j_{N}\right\rangle $, $\left\vert
i_{1}i_{2}...j_{p}...i_{N}\right\rangle $, and $\left\vert
j_{1}j_{2}...i_{p}...j_{N}\right\rangle $ is \ $\lambda ^{-}=-\left\vert
\det \left( \nu _{K}^{i_{1}i_{2}...i_{p}...i_{N}}\right) \right\vert $ with $%
\nu _{K}^{i_{1}i_{2}...i_{p}...i_{N}}$ defined as 
\begin{equation}
\nu _{K}^{i_{1}i_{2}...i_{p}...i_{N}}=\left[ 
\begin{array}{cc}
a_{i_{1}i_{2}...i_{p}...i_{N}} & a_{j_{1}j_{2}...i_{p}...j_{N}} \\ 
a_{i_{1}i_{2}...j_{p}...i_{N}} & a_{j_{1}j_{2}...j_{p}...j_{N}}%
\end{array}%
\right] ,  \label{nudef}
\end{equation}%
where $K=\sum\limits_{m=1}^{N}(1-\delta _{i_{m},j_{m}})$ $\left( 2\leq K\leq
N\right) $. In analogy with $\nu ^{00}$, $2\times 2$ matrix $\nu
_{K}^{i_{1}i_{2}...i_{p}...i_{N}}$ is defined as a $K-$way negativity font.
The subscript $K$ is used to group together the negativity fonts arising due
to $K-$way coherences of the composite system that is the correlations
responsible for GHZ state like entanglement of a $K-$partite system. For a
given value of $K$, the negativity of $K-$way partial transpose $\widehat{%
\rho }_{K}^{T_{p}}$ with respect to subsystem $p$, as defined in Ref. \cite%
{shar09}, arises solely from $K-$way negativity fonts. Determinants of
negativity fonts are, in a sense, intrinsic negative eigenvalues of a global
or a $K-$way partial transpose of state operator. Global partial transpose
of an $N-$qubit state is a combination of $K-$way partially transposed
operators $\left( 2\leq K\leq N\right) $ \cite{shar09} and can be expanded
as $\widehat{\rho }_{G}^{T_{p}}=\sum\limits_{K=2}^{N}\widehat{\rho }%
_{K}^{T_{p}}-(N-2)\widehat{\rho }$. Negativity of $\widehat{\rho }%
_{G}^{T_{p}}$, defined as $N_{G}^{A}=\left( \left\Vert \rho
_{G}^{T_{A}}\right\Vert _{1}-1\right) ,$where $\left\Vert \widehat{\rho }%
\right\Vert _{1}$ is the trace norm of $\widehat{\rho }$, arises due to all
possible negativity fonts present in $\widehat{\rho }_{G}^{T_{p}}$. Since $K$
qubits may be chosen in $\left( \frac{N!}{\left( N-K\right) !K!}\right) $
ways the form of a $K-$way font must specify the set of $K$ qubits it refers
to. To distinguish between different $K-$way negativity fonts we shall
replace subscript $K$ in Eq. (\ref{nudef}) by a list of qubit states for
which $\delta _{i_{m},j_{m}}=1$. In other words a $K-$way font involving
qubits $A_{1}$ to $A_{K}$ that is $\sum\limits_{m=1}^{N}(1-\delta
_{i_{m},j_{m}})=\sum\limits_{m=1}^{K}(1-\delta _{i_{m},j_{m}})=K$ reads as 
\begin{equation}
\nu _{\left( A_{K+1}\right) _{i_{K+1}}\left( A_{K+2}\right)
_{i_{K+2}}...\left( A_{N}\right) _{i_{N}}}^{i_{1}i_{2}...i_{p}...i_{N}}= 
\left[ 
\begin{array}{cc}
a_{i_{1}i_{2}...i_{p}...i_{K}i_{K+1}i_{K+2}...i_{N}} & 
a_{i_{1}+1i_{2}+1...i_{p}...i_{K}+1i_{K+1}i_{K+2},...i_{N}} \\ 
a_{i_{1}i_{2}...i_{p}+1...i_{K}i_{K+1}i_{K+2}...i_{N}} & 
a_{i_{1}+1i_{2}+1...i_{p}+1...i_{K}+1i_{K+1},i_{K+2}...i_{N}}%
\end{array}%
\right] ,  \label{nuk}
\end{equation}%
and its determinant is represented by 
\begin{equation}
D_{\left( A_{K+1}\right) _{i_{K+1}},\left( A_{K+2}\right)
_{i_{K+2}},...\left( A_{N}\right)
_{i_{N}}}^{i_{1}i_{2}...i_{p}...i_{K}}=\det \left( \nu _{\left(
A_{K+1}\right) _{i_{K+1}},\left( A_{K+2}\right) _{i_{K+2}},...\left(
A_{N}\right) _{i_{N}}}^{i_{1}i_{2}...i_{p}...i_{N}}\right) .  \label{dnuk}
\end{equation}%
Here $i_{m}+1=0$ for $i_{m}=1$ and $i_{m}+1=1$ for $i_{m}=0$. In this
notation no subscript is needed for an $N-$way negativity font that is $\nu
_{N}^{i_{1}i_{2}...i_{p}...i_{N}}=\nu ^{i_{1}i_{2}...i_{p}...i_{N}}.$

\section{Transformation of N-way Negativity fonts under local unitary on a
single qubit}

Determinant of an $N-$way negativity font

\begin{equation}
D^{i_{1}i_{2}...i_{p}=0...i_{N}}=\det \left[ 
\begin{array}{cc}
a_{i_{1}i_{2}...i_{p}=0...i_{N}} & a_{i_{1}+1,i_{2}+1,...i_{p}=0...i_{N}+1}
\\ 
a_{i_{1}i_{2}...i_{p}=1...i_{N}} & a_{i_{1}+1,i_{2}+1,...i_{p}=1...i_{N}+1}%
\end{array}%
\right] ,
\end{equation}%
is an invariant of local unitary U$^{A_{p}}$ acting on qubit $A_{p}$. After
applying unitary transformation $U^{A_{q}}=\frac{1}{\sqrt{1+\left\vert
x\right\vert ^{2}}}\left[ 
\begin{array}{cc}
1 & -x^{\ast } \\ 
x & 1%
\end{array}%
\right] $ on qubit $A_{q}$ with $q\neq p$ we obtain%
\begin{equation}
U^{A_{q}}\left\vert \Psi ^{A_{1},A_{2},...A_{N}}\right\rangle
=\sum_{i_{1}i_{2}...i_{N}}b_{i_{1}i_{2}...i_{N}}\left\vert
i_{1}i_{2}...i_{N}\right\rangle .  \label{uq}
\end{equation}%
Using primed symbols for determinants of negativity fonts calculated from
coefficients $b_{i_{1}i_{2}...i_{N}}$, we can write four transformation
equations%
\begin{eqnarray}
\left( D^{i_{1}i_{2}...i_{p}=0,i_{q}=0,...i_{N}}\right) ^{\prime } &=&\frac{1%
}{1+\left\vert x\right\vert ^{2}}\left[
D^{i_{1}i_{2}...i_{p}=0,i_{q}=0,...i_{N}}-\left\vert x\right\vert
^{2}D^{i_{1}i_{2}...i_{p}=0,i_{q}=1,...i_{N}}\right.  \notag \\
&&\left. +xD_{\left( A_{q}\right)
_{0}}^{i_{1}i_{2}...i_{p}=0...,i_{q-1},i_{q+1},...i_{N}}-x^{\ast }D_{\left(
A_{q}\right) _{1}}^{i_{1}i_{2}...i_{p}=0...,i_{q-1},i_{q+1},...i_{N}}\right]
\label{t1}
\end{eqnarray}%
\begin{eqnarray}
\left( D^{i_{1}i_{2}...i_{p}=0,i_{q}=1,...i_{N}}\right) ^{\prime } &=&\frac{1%
}{1+\left\vert x\right\vert ^{2}}\left[
D^{i_{1}i_{2}...i_{p}=0,i_{q}=1,...i_{N}}-\left\vert x\right\vert
^{2}D^{i_{1}i_{2}...i_{p}=0,i_{q}=0,...i_{N}}\right.  \notag \\
&&\left. +xD_{\left( A_{q}\right)
_{0}}^{i_{1}i_{2}...i_{p}=0...,i_{q-1},i_{q+1},...i_{N}}-x^{\ast }D_{\left(
A_{q}\right) _{1}}^{i_{1}i_{2}...i_{p}=0...,i_{q-1},i_{q+1},...i_{N}}\right]
\label{t2}
\end{eqnarray}%
\begin{eqnarray}
&&\left( D_{\left( A_{q}\right)
_{0}}^{i_{1}i_{2}...i_{p}=0...,i_{q-1},i_{q+1},...i_{N}}\right) ^{\prime }=%
\frac{1}{1+\left\vert x\right\vert ^{2}}\left[ x^{\ast }\left(
D^{i_{1}i_{2}...i_{p}=0,i_{q}=0,...i_{N}}+D^{i_{1}i_{2}...i_{p}=0,i_{q}=1,...i_{N}}\right) \right.
\notag \\
&&\left. +D_{\left( A_{q}\right)
_{0}}^{i_{1}i_{2}...i_{p}=0...,i_{q-1},i_{q+1},...i_{N}}+\left( x^{\ast
}\right) ^{2}D_{\left( A_{q}\right)
_{1}}^{i_{1}i_{2}...i_{p}=0...,i_{q-1},i_{q+1},...i_{N}}\right]  \label{t3}
\end{eqnarray}%
\begin{eqnarray}
&&\left( D_{\left( A_{q}\right)
_{1}}^{i_{1}i_{2}...i_{p}=0...,i_{q-1},i_{q+1},...i_{N}}\right) ^{\prime }=%
\frac{1}{1+\left\vert x\right\vert ^{2}}\left[ x\left(
D^{i_{1}i_{2}...i_{p}=0,i_{q}=0,...i_{N}}+D^{i_{1}i_{2}...i_{p}=0,i_{q}=1,...i_{N}}\right) \right.
\notag \\
&&\left. +D_{\left( A_{q}\right)
_{1}}^{i_{1}i_{2}...i_{p}=0...,i_{q-1},i_{q+1},...i_{N}}+x^{2}D_{\left(
A_{q}\right) _{0}}^{i_{1}i_{2}...i_{p}=0...,i_{q-1},i_{q+1},...i_{N}}\right]
\label{t4}
\end{eqnarray}%
relating $N-$way and $\left( N-1\right) -$way negativity fonts. Eliminating
variable $x$, invariants of $U^{A_{p}}U^{A_{q}}$ are found to be%
\begin{equation}
\left( D^{i_{1}i_{2}...i_{p}=0,i_{q}=0,...i_{N}}\right) ^{\prime }-\left(
D^{i_{1}i_{2}...i_{p}=0,i_{q}=1,...i_{N}}\right) ^{\prime
}=D^{i_{1}i_{2}...i_{p}=0i_{q}=0...i_{N}}-D^{i_{1}i_{2}...i_{p}=0i_{q}=1...i_{N}},
\label{didif}
\end{equation}%
\begin{eqnarray}
&&\left[ \left( D^{i_{1}i_{2}...i_{p}=0,i_{q}=0,...i_{N}}\right) ^{\prime
}+\left( D^{i_{1}i_{2}...i_{p}=0,i_{q}=1,...i_{N}}\right) ^{\prime }\right]
^{2}-4\left( D_{\left( A_{q}\right)
_{0}}^{i_{1}i_{2}...i_{p}=0...i_{N}}\right) ^{\prime }\left( D_{\left(
A_{q}\right) _{1}}^{i_{1}i_{2}...i_{p}=0...i_{N}}\right) ^{\prime }  \notag
\\
&=&\left(
D^{i_{1}i_{2}...i_{p}=0i_{q}=0...i_{N}}+D^{i_{1}i_{2}...i_{p}=0i_{q}=1...i_{N}}\right) ^{2}-4D_{\left( A_{q}\right) _{0}}^{i_{1}i_{2}...i_{p}=0...i_{N}}D_{\left( A_{q}\right) _{1}}^{i_{1}i_{2}...i_{p}=0...i_{N}}%
\text{,}  \label{disum}
\end{eqnarray}%
\begin{eqnarray}
&&\left( D^{i_{1}i_{2}...i_{p}=0,i_{q}=0,...i_{N}}\right) ^{\prime }\left(
D^{i_{1}i_{2}...i_{p}=0,i_{q}=1,...i_{N}}\right) ^{\prime }-\left( D_{\left(
A_{q}\right) _{0}}^{i_{1}i_{2}...i_{p}=0...i_{N}}\right) ^{\prime }\left(
D_{\left( A_{q}\right) _{1}}^{i_{1}i_{2}...i_{p}=0...i_{N}}\right) ^{\prime }
\notag \\
&=&D^{i_{1}i_{2}...i_{p}=0i_{q}=0...i_{N}}D^{i_{1}i_{2}...i_{p}=0i_{q}=1...i_{N}}-D_{\left( A_{q}\right) _{0}}^{i_{1}i_{2}...i_{p}=0...i_{N}}D_{\left( A_{q}\right) _{1}}^{i_{1}i_{2}...i_{p}=0...i_{N}}.
\label{diprod}
\end{eqnarray}%
Relevant multiqubit invariants for a given value of $N$ can be written down
from these general results. Invariants of $U^{A_{p}}U^{A_{r}}$ for $K-$way
fonts ($2\leq K\leq N$) with qubits $p$ and $r$ in the superscript and $N-K$
subscripts are analogous to those for $N-$way fonts.

\section{N-even N-way invariant}

Invariant of U$^{A_{1}}$U$^{A_{2}}$U$^{A_{3}}$ is obtained by taking a
combination of $N-$way invariants of U$^{A_{1}}$U$^{A_{2}}$ such that Eq. (%
\ref{didif}) is satisfied for the third qubit, for example 
\begin{equation}
I\left( U^{A_{1}}U^{A_{2}}U^{A_{3}}\right)
=D^{0000...0}-D^{0100...0}-D^{0010...0}+D^{0110...0}.
\end{equation}%
Using the same reasoning\ four qubit $N-$way invariant looks like%
\begin{eqnarray}
I\left( U^{A_{1}}U^{A_{2}}U^{A_{3}}U^{A_{4}}\right)
&=&D^{0000...0}-D^{0100...0}-D^{0010...0}+D^{0110...0}  \notag \\
&&-D^{0001...0}+D^{0101...0}+D^{0011...0}-D^{0111...0},
\end{eqnarray}%
and the $N-$way invariant for $N$ qubits reads as%
\begin{equation}
I_{N}=\sum\limits_{^{i_{2}..i_{N}}}\left( -1\right)
^{i_{1}+i_{2}...+i_{p}...+i_{N}}D^{0i_{2}...i_{N}}.
\end{equation}%
Noting that $D^{00i_{3}...i_{N}}=-D^{01i_{3}+1...i_{N}+1},$ we have%
\begin{equation}
\left( D^{00i_{3}...i_{N}}+\left( -1\right)
^{N-1}D^{01i_{3}+1...i_{N}+1}\right)
=D^{i_{1}i_{2}...i_{p}=0i_{q}=0...i_{N}}\left( 1+\left( -1\right)
^{N}\right) ,
\end{equation}%
\ giving $I_{N-odd}=0,$ while for N-even%
\begin{equation}
I_{N-even}=\sum\limits_{^{i_{3}...i_{N}}}\left( -1\right)
^{i_{3}+i_{4}+...+i_{N}}D^{00i_{3}...i_{N}}.  \label{in-even}
\end{equation}%
The invariant for $N-$even has permutation symmetry, as such may be used to
define $N-$tangle as 
\begin{equation}
\tau _{N-even}=4\left\vert
\sum\limits_{_{i_{1},i_{2},...i_{p-1},i_{p+1}...i_{q-1},i_{q+1},...,i_{N}}}%
\left( -1\right)
^{i_{1}+i_{2}...+i_{p}...+i_{N}}D^{i_{1}i_{2}...i_{p}=0i_{q}=0...i_{N}}%
\right\vert ^{2}.  \label{ntang}
\end{equation}%
Degree four invariants for $N$ qubits are obtained by starting with $N-2$
qubit $N$ way invariants and using Eq. (\ref{disum}) to obtain an $N$ qubit
invariant.

Four qubit $4-$way invariant with negativity fonts lying solely in $4-$way
partial transpose is written from Eq. ($\ref{in-even}$) as $%
I_{4}=D^{0000}+D^{0011}-D^{0010}-D^{0001}$. We identify $I_{4}$ with
invariant H of degree two as given in ref. \cite{luqu03}. A four qubit state
with four qubit entanglement arising due to quantum correlations of the type
present in a four qubit GHZ state, is distinguished from other entangled
states by a non zero $I_{4}$. This entanglement is lost without leaving any
residue, on the loss of a single qubit. The entanglement monotone based on $%
I_{4}$ is 
\begin{equation*}
\tau _{4}=4\left\vert \left[ D^{0000}+D^{0011}-\left(
D^{0010}+D^{0001}\right) \right] ^{2}\right\vert ,
\end{equation*}%
called four-tangle in analogy with three tangle \cite{coff00}. Four tangle $%
\tau _{4}$ vanishes on $W-$like state of four qubits, however, fails to
vanish on product of two qubit entangled states.

We now apply the method to construct entanglement monotones that detect the
entanglement of specific parts of the composite system, an entangled qubit
pair in this case. To obtain degree four invariants that detect products of
two qubit states, consider the combination of $4-$way fonts $%
J=D^{0000}-D^{0100}+D^{0010}-D^{0110}$, which is an invariant of U$^{A_{1}}$U%
$^{A_{2}}$. Using (Eq. (\ref{disum})), applied to four-way and three way
fonts the four qubit invariant is found to be%
\begin{eqnarray}
J^{A_{1}A_{2}} &=&\left( D^{0000}-D^{0100}+D^{0010}-D^{0110}\right) ^{2} 
\notag \\
&&+8D_{\left( A_{3}\right) _{0}\left( A_{4}\right) _{0}}^{00}D_{\left(
A_{3}\right) _{1}\left( A_{4}\right) _{1}}^{00}+8D_{\left( A_{3}\right)
_{1}\left( A_{4}\right) _{0}}^{00}D_{\left( A_{3}\right) _{0}\left(
A_{4}\right) _{1}}^{00}  \notag \\
&&-4\left( D_{\left( A_{3}\right) _{0}}^{000}-D_{\left( A_{3}\right)
_{0}}^{010}\right) \left( D_{\left( A_{3}\right) _{1}}^{000}-D_{\left(
A_{3}\right) _{1}}^{010}\right)  \notag \\
&&-4\left( D_{\left( A_{4}\right) _{0}}^{000}-D_{\left( A_{4}\right)
_{0}}^{010}\right) \left( D_{\left( A_{4}\right) _{1}}^{000}-D_{\left(
A_{4}\right) _{1}}^{010}\right) .  \label{j1}
\end{eqnarray}%
Similarly, invariant obtained by starting with four-way U$^{A_{1}}$U$%
^{A_{3}} $ invariant form is%
\begin{eqnarray}
J^{A_{1}A_{3}} &=&\left( D^{0000}-D^{0010}+D^{0001}-D^{0011}\right) ^{2} 
\notag \\
&&+8D_{\left( A_{2}\right) _{0}\left( A_{4}\right) _{0}}^{00}D_{\left(
A_{2}\right) _{1}\left( A_{4}\right) _{1}}^{00}+8D_{\left( A_{2}\right)
_{1}\left( A_{4}\right) _{0}}^{00}D_{\left( A_{2}\right) _{0}\left(
A_{4}\right) _{1}}^{00}  \notag \\
&&-4\left( D_{\left( A_{2}\right) _{0}}^{000}-D_{\left( A_{2}\right)
_{0}}^{010}\right) \left( D_{\left( A_{2}\right) _{1}}^{000}-D_{\left(
A_{2}\right) _{1}}^{010}\right)  \notag \\
&&-4\left( D_{\left( A_{4}\right) _{0}}^{000}-D_{\left( A_{4}\right)
_{0}}^{001}\right) \left( D_{\left( A_{4}\right) _{1}}^{000}-D_{\left(
A_{4}\right) _{1}}^{001}\right) ,  \label{j2}
\end{eqnarray}%
and starting with U$^{A_{1}}$U$^{A_{4}}$ invariant we get%
\begin{eqnarray}
J^{A_{1}A_{4}} &=&\left( D^{0000}-D^{0001}+D^{0010}-D^{0011}\right) ^{2} 
\notag \\
&&+8D_{\left( A_{2}\right) _{0}\left( A_{3}\right) _{0}}^{00}D_{\left(
A_{2}\right) _{1}\left( A_{3}\right) _{1}}^{00}+8D_{\left( A_{2}\right)
_{1}\left( A_{3}\right) _{0}}^{00}D_{\left( A_{2}\right) _{0}\left(
A_{3}\right) _{1}}^{00}  \notag \\
&&-4\left( D_{\left( A_{2}\right) _{0}}^{000}-D_{\left( A_{2}\right)
_{0}}^{001}\right) \left( D_{\left( A_{2}\right) _{1}}^{000}-D_{\left(
A_{2}\right) _{1}}^{001}\right)  \notag \\
&&-4\left( D_{\left( A_{3}\right) _{0}}^{000}-D_{\left( A_{3}\right)
_{0}}^{001}\right) \left( D_{\left( A_{3}\right) _{1}}^{000}-D_{\left(
A_{3}\right) _{1}}^{001}\right) ,  \label{j3}
\end{eqnarray}%
with corresponding entanglement monotones defined as $\beta ^{A_{1}A_{i}}=%
\frac{4}{3}\left\vert J^{A_{1}A_{i}}\right\vert $, $i=2-4$. By construction $%
\left\vert J^{A_{1}A_{i}}\right\vert $ detects entanglement between qubits $%
A_{1}A_{i}$, provided the pair $A_{1}A_{i}$ is entangled to its complement
in four qubit state. For qubit $A_{1}$ the invariants $J^{A_{1}A_{2}}$, $%
J^{A_{1}A_{3}}$, and $J^{A_{1}A_{4}}$ satisfy the relation $\left(
I_{4}\right) ^{2}=\frac{1}{3}\left(
J^{A_{1}A_{2}}+J^{A_{1}A_{3}}+J^{A_{1}A_{4}}\right) $. An interesting four
qubit state reported in Ref. \cite{yeo06} is%
\begin{equation}
\left\vert \chi \right\rangle =\frac{1}{\sqrt{8}}\left( \left\vert
0000\right\rangle +\left\vert 1111\right\rangle -\left\vert
0011\right\rangle +\left\vert 1100\right\rangle +\left\vert
1010\right\rangle -\left\vert 0101\right\rangle +\left\vert
0110\right\rangle +\left\vert 1001\right\rangle \right) .
\end{equation}%
which is known to have maximal entanglement of the pair $A_{1}A_{2}$ with
pair of qubits $A_{2}A_{4}$. The state can be rewritten as an entangled
state of $A_{1}A_{4}$ and $A_{2}A_{3}$ Bell pairs 
\begin{eqnarray*}
\left\vert \chi \right\rangle &=&\frac{1}{\sqrt{8}}\left( \left\vert
00\right\rangle _{A_{1}A_{4}}+\left\vert 11\right\rangle
_{A_{1}A_{4}}\right) \left( \left\vert 00\right\rangle
_{A_{2}A_{3}}+\left\vert 11\right\rangle _{A_{2}A_{3}}\right) \\
&&+\frac{1}{\sqrt{8}}\left( \left\vert 10\right\rangle
_{A_{1}A_{4}}-\left\vert 01\right\rangle _{A_{1}A_{4}}\right) \left(
\left\vert 10\right\rangle _{A_{2}A_{3}}+\left\vert 01\right\rangle
_{A_{2}A_{3}}\right)
\end{eqnarray*}
however, is not reducible to a pair of Bell states. We verify that for this
state $I_{4}=0$, $%
J^{A_{1}A_{2}}=J^{A_{1}A_{3}}=J^{A_{2}A_{4}}=J^{A_{3}A_{4}}=-\frac{1}{4}$,
and $J^{A_{1}A_{4}}=J^{A_{2}A_{3}}=\frac{1}{2}$. Therefore the state is
characterised by $\tau _{4}=0$, $\beta ^{A_{1}A_{2}}=\beta
^{A_{1}A_{3}}=\beta ^{A_{2}A_{4}}=\beta ^{A_{3}A_{4}}=\frac{1}{3}$, while $%
\beta ^{A_{1}A_{4}}=\beta ^{A_{2}A_{3}}=\frac{2}{3}$, indicating that the
entanglement of state $\left\vert \chi \right\rangle $ is distinct from that
of GHZ state of four qubits having $\tau _{4}=1$, $\beta ^{A_{1}A_{2}}=\beta
^{A_{1}A_{3}}=\beta ^{A_{1}A_{4}}=\frac{1}{3}$ as well as $\beta
^{A_{2}A_{3}}=\beta ^{A_{2}A_{4}}=\beta ^{A_{3}A_{4}}=\frac{1}{3}$.

The degree four invariants for four qubits denoted as $L$, $M$, and $N$ in
Ref. \cite{luqu06} are combinations of $J^{A_{1}A_{2}}$, $J^{A_{1}A_{3}}$, $%
J^{A_{1}A_{4}}$ and $\left( I_{4}\right) ^{2}$. Additional invariants are
easily constructed to detect all possible types of Four qubit entanglement.\
One can verify that different types of four qubit entanglement detected by
antilinear operators of ref. \cite{oste05} are quantified by entanglement
monotones constructed from four qubit invariants.

\section{N-odd N-way invariant}

Since $I_{N-odd}=0$, there is no degree two invariant of $N-$way fonts for a
general state of $N-$odd qubits. But we can single out a qubit, write $N-1$
qubit invariants and then use Eq. (\ref{disum}) to obtain $N-$qubit
invariant. If we single out $N^{th}$ qubit and look at negativity fonts of $%
\rho ^{T_{A_{1}}}$, then two $\left( N-1\right) $ qubit $N-$way invariants
are%
\begin{equation}
I_{N-way}^{A_{1}\left( A_{N}\right)
_{0}}=\sum\limits_{i_{3}...,i_{N-1}}\left( -1\right)
^{i_{3}+...+i_{N-1}}D^{00i_{3}...i_{N-1}i_{N}=0},
\end{equation}%
\begin{equation}
I_{N-way}^{A_{1}\left( A_{N}\right)
_{1}}=\sum\limits_{i_{3}...,i_{N-1}}\left( -1\right)
^{i_{3}+...+i_{N-1}}D^{00i_{3}...i_{N-1}i_{N}=1}.  \label{inzero}
\end{equation}%
Transformation equations for $I_{N-way}^{A_{1}\left( A_{N}\right) _{0}}$ and 
$I_{N-way}^{A_{1}\left( A_{N}\right) _{1}}$, under unitary U$^{A_{N}}$ are
written by using Eqs. (\ref{t1}) to (\ref{t4}) and yield an $N-$qubit
invariant

\begin{equation}
I_{N-odd}^{A_{1}A_{N}}=\left( I_{N-way}^{A_{1}\left( A_{N}\right)
_{0}}+I_{N-way}^{A_{1}\left( A_{N}\right) _{1}}\right) ^{2}-4I_{\left(
N-1\right) -way}^{A_{1}\left( A_{N}\right) _{0}}\times I_{\left( N-1\right)
-way}^{A_{1}\left( A_{N}\right) _{1}}  \label{inodd}
\end{equation}%
with negativity fonts in $\rho ^{T_{A_{1}}}$, where%
\begin{equation}
I_{\left( N-1\right) -way}^{A_{1}\left( A_{N}\right)
_{0}}=\sum\limits_{i_{3}...,i_{N-1}}\left( -1\right)
^{i_{3}+...+i_{N-1}}D_{\left( A_{N}\right) _{0}}^{00i_{3}...i_{N-1}},
\label{inmoneway}
\end{equation}%
and%
\begin{equation}
I_{\left( N-1\right) -way}^{A_{1}\left( A_{N}\right)
_{1}}=\sum\limits_{i_{3}...,i_{N-1}}\left( -1\right)
^{i_{3}+...+i_{N-1}}D_{\left( A_{N}\right) _{1}}^{00i_{3}...i_{N-1}},
\label{in1nmneway}
\end{equation}%
are $\left( N-1\right) -$way invariants of local unitaries on $\left(
N-1\right) $ qubits (even no of qubits). Similarly, one may construct $\tau
_{N-odd}^{A_{1}A_{p}}$ for $2\leq p\leq N$ and $N+1\rightarrow 1$ (mod N).
Entanglement monotone based on $I_{N-odd}^{A_{1}A_{p}}$ is $\tau
_{N-odd}^{A_{1}A_{p}}=4\left\vert I_{N-odd}^{A_{1}Ap}\right\vert $. 
For $N=3$, three qubit invariant of degree two determines three-tangle \cite%
{coff00}, $\tau _{3}=4\left\vert \left( D^{000}-D^{001}\right)
^{2}-4D_{\left( A_{2}\right) _{0}}^{00}D_{\left( A_{2}\right)
_{1}}^{00}\right\vert $. For $N=5$, the five-way invariants of local
unitaries on qubits $A_{1}$, $A_{2}$, $A_{3}$, and $A_{4}$ and corresponding
four-way invariants combine to give
 
\begin{eqnarray}
I_{5}^{A_{1}A_{5}} &=&\left(
D^{00000}-D^{00010}+D^{00110}-D^{00100}+D^{00001}-D^{00011}+D^{00111}-D^{00101}\right) ^{2}
\notag \\
&&-4\left( D_{\left( A_{5}\right) _{0}}^{0000}-D_{\left( A_{5}\right)
_{0}}^{0001}-D_{\left( A_{5}\right) _{0}}^{0010}+D_{\left( A_{5}\right)
_{0}}^{0011}\right) \left( D_{\left( A_{5}\right) _{1}}^{0000}-D_{\left(
A_{5}\right) _{1}}^{0001}-D_{\left( A_{5}\right) _{1}}^{0010}+D_{\left(
A_{5}\right) _{1}}^{0011}\right)  \label{ifive}
\end{eqnarray}%
which is a five qubit invariant of degree four with fonts in five way and
four-way partial transpose with respect to qubit $A_{1}$. In general, one
can construct $I_{5}^{A_{p}A_{q}}$ obtaining a five tangle $\tau
_{5}^{A_{p}A_{q}}=4\left\vert I_{5}^{A_{p}A_{q}}\right\vert $for each choice
of p and q value. Degree four invariants to detect entanglement of two
entangled qubits with their compliment in a five qubit state are
combinations of two qubit invariants of five way, four-way, three way and
two way fonts and can be obtained in a way analogous to that for five tangle.

To conclude, local unitary polynomial invariants for $N$ qubit quantum state
have been obtained from basic units of entanglement, referred to as
negativity fonts.\ The method exploits the transformation properties of
determinants of $K-$way negativity fonts under local unitary
transformations. The entanglement monotone based on square of degree two
invariant\ for $N$ even (Eq. (\ref{ntang})) and degree four invariant of Eq.
(\ref{inodd}) for $N-$odd is referred to as $N-$tangle in analogy with
three-tangle \cite{coff00}. The method aims at obtaining LU invariants, that
are relevant to classifying multi-qubit entangled states. To illustrate the
construction of entanglement monotones that detect entanglement of specific
parts of the composite system, degree four invariants to detect entanglement
of entangled pairs in a four qubit state are reported. Our method can be
used to generate the relevant invariants obtained by using different
approaches in references \cite{wong01,luqu03,luqu06,luqu07,li07} and also to
generate additional invariants necessary to detect specific entanglement
modes. Entanglement monotones constructed from invariants can identify the
class to which a given state belongs. Local unitary transformations
redistribute the negativity fonts amongst $K-$way partial transposes and may
also reduce the number of negativity fonts in a given partial transpose. To
determine unitary transformations that relate two unitary equivalent states
is an important question in quantum information. The key to determine the
unitary transformations relating two states belonging to the same class lies
in the numerical value of invariants, number and type of negativity fonts
and transformation equations that the determinants of negativity fonts for
each state satisfy. The transformation equations for negativity fonts can be
used directly to identify the unitaries that relate two unitary equivalent
states. The method can be easily extended to qutrits and higher-dimensional
systems.

Financial support from CNPq, Brazil and Funda\c{c}\~{a}o Arauc\'{a}ria,
Brazil is acknowledged.

\end{document}